\documentclass[11pt,notoc]{JHEP3}

\usepackage{graphicx}

\usepackage{fancybox}
\usepackage{epsfig}
\usepackage{dcolumn}
\usepackage{bm}
\usepackage{amsfonts}
\usepackage{mathrsfs}
\usepackage{amssymb}
\usepackage{amsmath}

\def\al{\alpha}
\def\be{\beta}

\def\ep{\epsilon}

\def\ka{\kappa}

\newcommand{\ben}{\begin{equation}}
\newcommand{\een}{\end{equation}}
\newcommand{\bea}{\begin{eqnarray}}
\newcommand{\eea}{\end{eqnarray}}
\newcommand{\ba}{\begin{array}}
\newcommand{\ea}{\end{array}}
\newcommand{\bit}{\begin{itemize}}
\newcommand{\eit}{\end{itemize}}

\title{Inhomogeneous tachyon condensation}

\author{Mark Hindmarsh and Huiquan Li \\
Department of Physics and Astronomy, \\
University of Sussex, \\
Brighton BN1 9QH, \\
U.K.}

\abstract{We investigate the spacetime-dependent condensation of
the tachyon in effective field theories. Previous work identified
singularities in the field which appear in finite time: infinite
gradients at the kinks, and (in the eikonal approximation)
caustics near local minima.
By performing a perturbation analysis, and with numerical
simulations, we demonstrate and explain key features of the
condensation process: perturbations generically freeze, and
minima develop singular second derivatives in finite time
(caustics). This last has previously been understood in terms of
the eikonal approximation to the dynamics. We show explicitly from
the field equations how this approximation emerges, and how the 
caustics develop, both in the DBI and BSFT effective actions.
We also investigate the equation of state parameter of tachyon 
matter showing that it is small, but generically non-zero. The
energy density tends to infinity
near field minima with a charateristic profile. A proposal to
regulate infinities by modifying the effective action is also
studied. We find that although the infinities at the kinks are
successfully regularised in the time-dependent case, caustics
still present.}


\begin{document}

\section{Introduction}
\label{sec:Introduction}

In Type II string theories, unstable branes of dimension $p$
decaying into the stable ones with lower dimensions is well
described by the Dirac-Born-Infeld (DBI) type effective action
\cite{Garousi:2000tr,Bergshoeff:2000dq,Sen:2002an}:
\begin{equation}
\label{E:DBIA}
 S=-\int d^{p+1} x V(T)\sqrt{1+y},
\end{equation}
where $y=\eta^{\mu\nu}\partial_\mu{T}\partial_\nu{T}$, $V(T)$ is
the asymptotic potential, and we use units in which $\al'=1$. At
the beginning of the condensation, the system has zero tachyon
field $T=0$ and is located on the top of the potential $V(T)$. It
is unstable and will roll down when driven by a small
perturbation. The field evolves towards the minimum of the
potential $V=0$, and the condensation ends. The final state
corresponds to the disappearance of the unstable branes, replaced
by ``tachyon matter" whose properties are not well-understood.

The condensation process in the inhomogeneous case has also been
studied. It has been realised that the equation of motion from the
DBI action leads to solitonic solution with kinks and anti-kinks
\cite{Hashimoto:2002ct,Brax:2003rs,Kim:2003in,Hindmarsh:2007dz}.
At the kinks and anti-kinks, the field remains zero, and become
daughter branes of one dimension lower at the end of the
condensation. While the field in between them grows with time and
tends to infinity. The huge difference of the field values between
these two areas makes the field gradient at kinks and anti-kinks
very large and in fact can be semi-analytically shown to lead to
infinity in finite time \cite{Cline:2003vc,Hindmarsh:2007dz}. The
existence of this kind of singularity prevents straightforward
numerical integration of the space-time dependent field equation
\cite{Felder:2002sv,Barnaby:2004dz}.

In between the kinks and anti-kinks where the field is approximately
homogeneous, an eikonal approximation $1+y=0$ reveals that the 
second and higher derivatives of the field can reach infinity in 
finite time. This is interpreted as the formation of caustics
\cite{Felder:2002sv}.

The potential application of the tachyon condensation in the
inflation scenario has been investigated, regardless of the
various problems
\cite{Mazumdar:2001mm,Fairbairn:2002yp,Choudhury:2002xu,
Kofman:2002rh,Sami:2002fs,Shiu:2002qe}. In the tachyon inflation
model, agreement with the observational data can be achieved
\cite{Steer:2003yu}. A complex tachyon field can produce cosmic
strings after inflation
\cite{Dvali:1998pa,Sarangi:2002yt,Choudhury:2003vr}. The
appearance of the pressureless tachyon matter at late stage of the
condensation \cite{Sen:2002in,Sen:2002an} is also a feature of
this model. The fluctuations of the tachyon matter in both the
free and the expanding cosmology background
\cite{Frolov:2002rr,Kofman:2002rh,Shiu:2002qe,Hashimoto:2002ct,
Steer:2003yu} have been studied, and shown to increase linearly
with time, just like ordinary pressureless matter.

In this paper, we study the condensationg process using the 
field equations, demonstrating and explaining, both analytically
and numerically, key features of the inhomogeneous tachyon
condensation process. Perturbations
around the homogeneous solutions $T=t$ ``freeze'', that is, show
little change after a short initial relaxation period. What change
there is reduces the curvature at maxima, and increases it at minima,
until it becomes singular in finite time. This is interpreted as the
formation of a caustic in the eikonal solution.

We also show explicitly from the field equations, backed up with
numerical simulations, that the quantity $1+y$ relaxes exponentially
to zero, thus accounting for the accuracy of the eikonal equation.

We also investigate the energy density and pressure of condensing
tachyon matter,
showing that it is not quite pressureless, and that the energy
density diverges in a characteristic $1/|\triangle x|$ from near
the developing caustic.

One can ask if the instability is an artifact of the DBI effective
action. To give a partial answer, we also study the boundary
string field theory (BSFT) effective action
\cite{Kutasov:2000aq,Kraus:2000nj}, showing that this approaches
the eikonal equation eventually.

Finally, we show that a modification of the action proposed by
\cite{Brax:2003rs} to regulate the gradient at a static kink does
prevent the gradient of the field reaching infinity at the kinks
in the time-dependent case, but does not prevent caustic formation
between kinks. 

The paper is constructed as follows. In Sec.\ 2, we give the
equation of motion and its solutions in simple cases. In Sec.\ 3,
we present the $1+1$ dimensional solutions near kinks and near
extrema respectively. In Sec.\ 4 and Sec.\ 5, we discuss the
dynamics approaching the vacuum respectively in the DBI and the
BSFT effective theories. Based on the results, the features of 
the tachyon matter towards the end of the condensation are
investigated in Sec.\ 6. In Sec.\ 7, we discuss the generalised
DBI action.

\section{Equation of motion}
\label{sec:EoM}

From the DBI action (\ref{E:DBIA}), we can write down the equation
of motion:
\begin{equation}\label{e:feq1}
 \left(\Box T-\frac{V'}{V}\right)(1+y)
=\frac{1}{2}\partial^\mu T\partial_\mu(1+y),
\end{equation}
where $\Box=\eta_{\mu\nu}\partial^\mu\partial^\nu$.
An equivalent expression to the equation of motion is
\begin{equation}
\label{E:EoM}
 \ddot{T}=f\left\{2\dot{T}\nabla_i\dot{T}\nabla^iT+(1+y)
\nabla^2T-\nabla_i T\nabla_j T\nabla^i \nabla^jT-\frac{V'}
{V}(1+y)\right\},
\end{equation}
where $f=1/(1+\nabla_i{T}\nabla^i{T})$,
$\nabla^2=\partial_i\partial^i$ and $i$, $j=1,\cdots,p$. $V(T)$ is
the runaway tachyon potential and its field derivative is
$V'=dV(T)/dT$. A suitable choice of potential, derived from the
boundary conformal field theory on the worldsheet
\cite{Kutasov:2003er,Smedback:2003ur}, is
\begin{equation}
\label{E:Pot1}
 V=\frac{V_m}{\cosh{(\beta T)}},
\end{equation}
where the constant $\beta=1$ for the bosonic string and
$\beta=1/\sqrt{2}$ for superstring.

The energy-momentum tensor is:
\begin{equation}\label{E:enemom}
 T_{\mu\nu}=V(T)\left[\frac{\partial_\mu T\partial_\nu T}
{\sqrt{1+\partial T\cdot\partial T}}-\eta_{\mu\nu}\sqrt
{1+\partial T\cdot\partial T}\right].
\end{equation}

\subsection{The static case}

For the 1-dimensional static case, the solution to the field
equation is most easily found by noting that conservation of
energy-momentum requires that the pressure is constant:
\begin{equation}
T_{11}=\frac{-V}{\sqrt{1+{T'}^2}}=-V_0,
\end{equation}
where $V_0=V(T_0)$ is the minimum potential when the maximum field
$T_0$ is achieved at $x=x_0$. The maximum field satisfies the
condition $T'|_{T=T_0}=0$.

For the inverse hyperbolic potential (\ref{E:Pot1}), the equation
is solvable:
\begin{equation}
\label{E:Tsol}
 T(x)=\frac{1}{\beta}\sinh^{-1}{\left[\sqrt{\frac{1}{r^2}-1}
\sin(\beta\triangle x)\right]},
\end{equation}
where $r=V_0/V_m$ and $\triangle x=x-x_m$. It represents an array
of solitonic kinks and anti-kinks. Its gradient is:
\begin{equation}
 T'=\pm\sqrt{\frac{1-r^2}{r^2+\tan^2{(\beta \triangle x)}}}.
\end{equation}
For fixed $V_m$, $T'$ tends to infinity at kinks and anti-kinks,
where $\triangle x=n\pi/\beta$, as $V_0\rightarrow 0$, which is
consistent with the slow motion analysis on the moduli space of
\cite{Hindmarsh:2007dz}.

\subsection{The homogeneous case}

In this case, the energy density is conserved and constant:
\begin{equation}
T_{00}=\frac{V}{\sqrt{1-\dot{T}^2}}=E,
\end{equation}
which restricts $|\dot{T}|\leq 1$.

For the potential (\ref{E:Pot1}), the time dependent solution
is:
\begin{equation}
 T(t)=
 \left\{
 \begin{array}{cl}
 \frac{1}{\beta}\sinh^{-1}[\sqrt{\frac{1}{l^2}-1}
\cosh{(\beta t)}], & (E\leq V_m), \\
 \frac{1}{\beta}\sinh^{-1}[\sqrt{1-\frac{1}{l^2}}
\sinh{(\beta t)}], & (E>V_m),
 \end{array}
 \right.
\end{equation}
and its time derivative is:
\begin{equation}
 \dot{T}=
 \left\{
 \begin{array}{cl}
 \pm\sqrt{\frac{l^2-1}{l^2-\tanh^2{(\beta t)}}}, & (E\leq V_m), \\
 \pm\sqrt{\frac{1-l^2}{\coth^2(\beta t)-l^2}}, & (E>V_m),
 \end{array}
 \right.
\end{equation}
where $l=E/V_m$. For $E>V_m$, $T=0$, $|\dot{T}|=\sqrt{1-1/l^2}$
at $t=0$ and $|T|\rightarrow\infty$,
$|\dot{T}|\rightarrow 1$ as $t\rightarrow\infty$.

\section{Solutions near kinks and extrema}
\label{sec:solutions}

In this section, we study the behaviour of solutions near
stationary kinks and extrema, in order to gain insight into the
singular dynamics, and to make contact with previous work
\cite{Felder:2002sv,Cline:2003vc}.

\subsection{Around kinks}

Following \cite{Cline:2003vc}, we expand the field around a
stationary kink or anti-kink located at $x=x_m$:
\begin{equation}
\label{E:Exp1}
 T(t,x) = a(t)\triangle x+\frac{1}{6}c(t)\triangle x^3+
\frac{1}{120}e(t)\triangle x^5+\cdots,
\end{equation}
where $\triangle x=x-x_m$.

For the potential (\ref{E:Pot2}), $V'/V=-\beta^2 T$. For the
potential (\ref{E:Pot1}), $V'/V \simeq -\beta^2 T$ when $|T|$ is
small. We use this relation and the expansion (\ref{E:Exp1}) in
the field equation around kinks and anti-kinks, where $|T|$ is
small. The comparison of coefficients gives the equation:
\begin{equation}
 \ddot{a} = \beta^2 a+\frac{2a\dot{a}^2+c}{1+a^2}.
\end{equation}

We are more interested in the solution at late time when $a(t)$
becomes large. If we can neglect the contribution from $c(t)$,
\begin{equation}
\label{E:at}
 \ddot{a}a = \beta^2 a^2+2\dot{a}^2.
\end{equation}
To solve the equation, we set $a=1/z$, giving
\begin{equation}
\label{E:zt}
 \ddot{z} + \beta^2 z=0,
\end{equation}
with solution
\begin{equation}
 z=z_1 \cos{(\beta t)}+z_2 \sin{(\beta t)}.
\end{equation}
By a suitable choice of the time coordinate, the solution of $a$
can be written
\begin{equation}
\label{E:aSol}
 a(t)=\frac{a_0}{\cos{(\beta t)}},
\end{equation}
where $a_0$ is constant. If we set $t_*=\pi/(2\beta)$, we
approximately have: $a\sim 1/(t_*-t)$, which is consistent with
the result of \cite{Cline:2003vc}. $a(t)$ grows to infinity in a
finite time.

More precisely, we now also consider the evolution of the
coefficient $c(t)$. With the same approximation of $a$ being
large, we can obtain the equation:
\begin{equation}
 a^2\ddot{c}=8a\dot{a}\dot{c}+(7\beta^2a^2-6a\ddot{a})c
-6\beta^2a\dot{a}^2+e.
\end{equation}
Combining with Eq.\ (\ref{E:at}) and neglecting $e(t)$, we find
\begin{equation}
 c(t)=-\beta^2 a=-\frac{\beta^2 a_0}{\cos{(\beta t)}}.
\end{equation}
Therefore, the solution near the kinks and anti-kinks can be
expressed as:
\begin{equation}
\label{E:KinT}
 T(t,x)=a(t)\left[\triangle x-\frac{1}{6}\beta^2{\triangle x}^3
+\cdots\right]\simeq{\frac{a_0}{\beta}\frac{\sin{(\beta\triangle x)}}
{\cos{(\beta t)}}}.
\end{equation}
This has the following properties at $x=x_m$: $T=0$, $\dot{T}=0$,
$T'(t\rightarrow\infty)\rightarrow \infty$, $\ddot{T}=0$ and
$T''=0$, which are consistent with the discussion in Sec.\ 2.

\subsection{Around extrema}

The field around a stationary peak or trough at $x=x_0$ satisfies
the condition $T'|_{x=x_0}=0$. The expansion around $x_0$ is
written:
\begin{equation}
\label{E:Exp2}
 T(t,x) = T_0(t)+\frac{1}{2}b(t)\triangle x^2+\frac{1}{24}d(t)
\triangle x^4+\cdots,
\end{equation}
with $\triangle x=x-x_0$. For convenience, we only consider the
behaviour around the field extrema whose values are positive
$T_0(t)>0$.

For the potential (\ref{E:Pot1}), $V'/V\simeq -\beta$ when $T$ is
positive. Inserting the expansion form in the field equation
(\ref{E:EoM}), we get:
\begin{equation}\label{E:pi0eq1}
 \ddot{T}_0 = (\beta+b)(1-\dot{T}_0^2).
\end{equation}
The final value of $\dot{T}_0$ is decided by the relation between
$b(t)$ and $\beta$. Here we only consider two special cases. When
$b=-\beta$, $\dot{T}_0$ is constant and it corresponds to the
static case. This is consistent to the static analytical solution
(\ref{E:Tsol}), which has $T''=-\beta$ at peaks. Secondly, if
$|b|\ll\beta$, there is an explicit solution for $\dot{T}_0$
\begin{equation}
\label{E:pi0}
 \dot{T}_0=\tanh(\be t).
\end{equation}
The solution satisfies $\dot{T}_0\in[0,1]$. Otherwise, the
solution can be formally expressed as:
$\dot{T}_0\simeq1-(A/2)e^{-2(\be+b)t}$, where $A$ is a small
positive value.

The equation for the coefficient $b(t)$ is:
\begin{equation}\label{E:beq1}
 \ddot{b}=2\beta b^2-2(\beta-b)\dot{T}_0\dot{b}.
\end{equation}
Since we are considering the peaks on the positive field side
$T_0>0$, $\dot{T}_0$ should be positive and tends to 1. Making
use of the approximation $\dot{T}_0 \simeq 1$ for late time,
Eq.\ (\ref{E:beq1}) can be rewritten as:
\begin{equation}
\label{E:bt1}
 \partial_t{(\dot{b}-b^2)} = -2\beta(\dot{b}-b^2).
\end{equation}
Hence,
\begin{equation}
\label{E:bSol1}
 \dot{b} = b^2 + C\exp{(-2\beta t)},
\end{equation}
where $C=\dot{b}_0-b_0^2$ is a constant, with $b_0=b(t=0)$ and
$\dot{b}_0=(db/dt)|_{t=0}$. We have made numerical solutions of
(\ref{E:bSol1}). The distribution of the final values as a
function of $b_0$ and $\dot{b}_0$ is presented in Fig.\
\ref{f:bdis}. From these results, we see that a peak ($b_0<0$) is
most likely to flatten with time, but a trough ($b_0>1$) can
deepen and become singular, and will always do so when $\dot{b}_0$
is positive. There is a sign of the developement of a caustic.

\FIGURE[ht]{
\epsfig{file=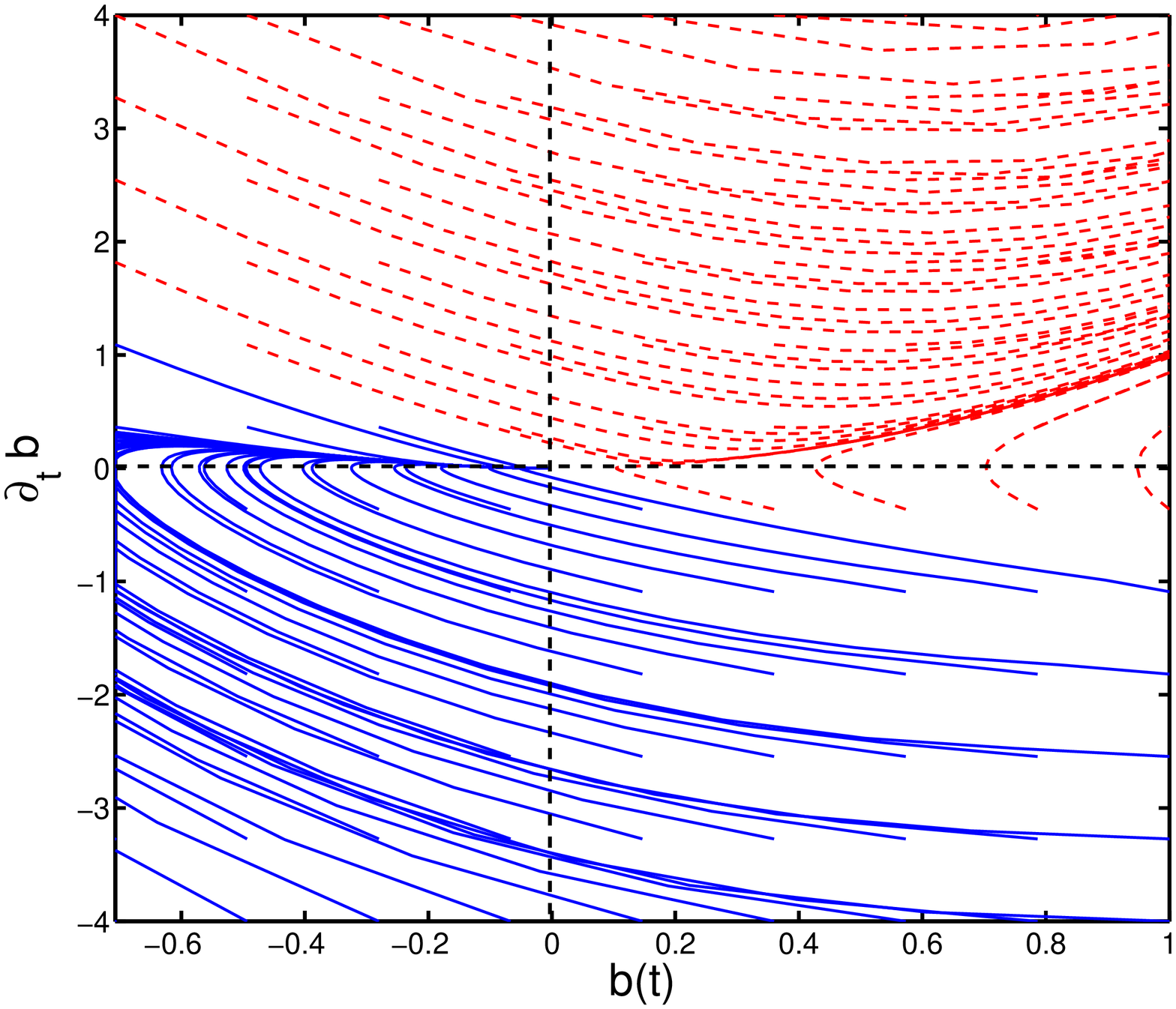,scale=0.35}
\caption{ The time evolving curves of $\dot{b}(t)$ vs $b(t)$ with
different initial values ($b_0$, $\dot{b}_0$) in regions where
$b\geq -\be$. The blue, solid lines indicate that $b(t)\rightarrow
0$ at large $t$ and the red, dashed lines indicate that $b(t)$
tends to infinity. \label{f:bdis}} }

Focus on the late time behaviour of $b(t)$. In terms of Eq.\
(\ref{E:bt1}), we will have as $t\rightarrow\infty$
\begin{equation}
 \dot{b} = b^2.
\end{equation}
So we can have the solution of $b$
\begin{equation}
 b(t)=\frac{b_0}{1-b_0t},
\end{equation}
where $b_0=b(t=0)$. Thus $b$ blows up at $t=t_0$ when $b_0>0$
while $b\rightarrow0$ when $b_0\leq0$. But $\dot{b}\geq0$ for both
cases. This is consistent with the numerical solutions prsented in
Fig.\ (\ref{f:bdis}).

\subsection{Caustics around extrema}

In terms of the relations in Eq.\ (\ref{E:pi0}) and
(\ref{E:bSol1}), we can estimate the quantity $1+y$ at late time
as:
\begin{equation}\label{E:Rel}
 1+y\simeq1-\dot{T}_0^2-(\dot{b}-b^2)\triangle x^2
\simeq (Ae^{-2bt}-C\triangle x^2)e^{-2\beta t},
\end{equation}
If $b>-\be$, we see that $1+y\rightarrow 0$, first noted in
\cite{Felder:2002sv}. If we are in the region $\dot{b}<b^2$,
$C$ is negative and $1+y$ approaches zero from above.

Based on this relation, the authors of \cite{Felder:2002sv} argue
that there should be caustics forming around the peaks or troughs.
They rewrite the relation (\ref{E:Rel}) as an eikonal
equation:
\begin{equation}\label{E:HJ}
 \dot{S}^2-S'^2=1,
\end{equation}
with the approximation $S\simeq T$ and develop a series of
characteristic equations for (\ref{E:HJ}) to give
\begin{equation}
 S''(t,x)=\frac{S''(0,x)}{1-\frac{S''(0,x)}{(1+S'^2(0,x))^{3/2}}t}.
\end{equation}
Thus $S''$ blows up in a finite time
$t=(1+S'^2(0,x))^{3/2}/S''(0,x)$ when $S''(0,x)>0$, which is
interpreted as the appearance of caustics or regions where the
tachyon field becomes multivalued. We also note that when
$S''(0,x)\leq0$ there are no caustics, with $S''(t,x)\rightarrow0$
as $t\rightarrow\infty$.

Our results are consistent with the HJ analysis above:
caustics form near troughs ($b>0$) in a time $b_0^{-1}$, where
$b_0\simeq S''(0,x)$ at a trough, and peaks tend to flatten
($b\rightarrow0$).

\section{Perturbations}
\label{sec:perturbation}

In this section, we will consider the evolution of perturbations
around a condensing homogeneous tachyon field in an arbitrary
dimensional D$p$-brane. This will also approximately describes the
evolution of the field between well-separated kinks and
anti-kinks. A general form is:
\begin{equation}
\label{E:Pert}
 T(t,x^i) = T_0(t)+\tau(t,x^i),
\end{equation}
where $\tau(t,x^i)$ is the small perturbation along all spatial
directions along the D$p$-brane $x^i$ with $i=1,\cdots,p$.

A trivial case is: $T_0(t)=0$. Around $T_0(t)=0$, $-V'/V\simeq
\beta^2 \tau$ for both potentials (\ref{E:Pot1}) and
(\ref{E:Pot2}). Inserting $T(t,x^i)=\tau(t,x^i)$ in the field
equation (\ref{E:EoM}), we can get the perturbation equation:
\begin{equation}
 \nabla^2\tau-\ddot{\tau}+\beta^2\tau=0,
\end{equation}
as expected for a tachyon field with mass squared $-\beta^2$.

\subsection{Linear perturbations around a condensing field}

Plugging the ansatz (\ref{E:Pert}) in the field equation with the
hyperbolic potential (\ref{E:Pot1}), we have the equation of
$\tau$
\begin{eqnarray}\label{E:pereq}
 [\ddot{T}_0-\beta(1-\dot{T}_0^2)]+[\ddot{\tau}+2\beta
\dot{T}_0\dot{\tau}-(1-\dot{T}_0^2)\nabla^2\tau]+,
\\ \nonumber
[\be\dot{\tau}^2-(\be-\ddot{T}_0)\nabla\tau\cdot
\nabla{\tau}-2\dot{T}_0(\nabla\tau\cdot\nabla
\dot{\tau}-\dot{\tau}\nabla^2\tau)]
\\ \nonumber
-[(\partial\tau\cdot\partial\tau)\Box\tau-\partial_\mu
\tau\partial_\nu\tau\partial^\mu\partial^\nu\tau]=0.
\end{eqnarray}
Solving the zeroth order equation covered in the first square
bracket, we get the solution $\dot{T}_0=\tanh(\be t)$, as given in
Eq.\ (\ref{E:pi0}).

We first consider the linear perturbation covered in the second
square bracket. This situation applies if all the first order
terms remains dominated compared to the second order terms, though
this is always not true as we can see later.

The solution of $\tau$ in the second square bracket is separate,
and writing $\tau(t,x)=f(t)g(x)$, and we find:
\begin{equation}
\label{e:2Eq}
 \left\{
 \begin{array}{c}
 \nabla^2 g+k^2 g=0, \\
 \ddot{f}+2\be\tanh(\be t)\dot{f}+k^2\mathrm{sech}^2
(\be t)f=0,
 \end{array}
 \right.
\end{equation}
where $k^2=k_ik^i$ and $k^i$ are arbitrary constants.

By making the reparameterisation
\begin{eqnarray}
 \rho=\frac{1}{1+e^{-2\be t}}, & (\frac{1}{2}\leq\rho\leq 1),
\end{eqnarray}
we can rewrite the second equation in Eq.\ (\ref{e:2Eq}) as
\begin{equation}
 \rho(1-\rho)\frac{d^2f}{d\rho^2}+\ka^2f=0.
\end{equation}
where $\ka^2=\ka_i\ka^i$ and $\ka_i=k_i/\be$. It is a typical
hypergeometric differential equation. The equation has two
linearly independent solutions, the simpler one of which is given
as \cite{Jeffrey:2007}
\begin{equation}\label{e:hypergsol}
 f(t)=\rho(t)_2F_1\left(\frac{1-\sqrt{1+4\ka^2}}{2},\frac{1+
\sqrt{1+4\ka^2}}{2};2;\rho(t)\right),
\end{equation}
where $_2F_1$ is the hypergeometric function.

We will focus on the late time behaviour of $\tau$ when
$\dot{T}_0$ is closed to 1. In this situation, the second equation
in Eq.\ (\ref{e:2Eq}) becomes
$\ddot{f}+2\beta\dot{f}+4k^2\exp{(-2\beta t)}f=0$. With the
reparameterization $\eta=\exp{(-2\beta t)}$, we have:
\begin{equation}
 \eta\frac{d^2 f}{d\eta^2}+\ka^2f=0.
\end{equation}
The equation is a Bessel style differential equation
\cite{abramowitz+stegun}. Its solution is given as a conbination
of two Bessel functions. The whole solution at late time can be
expressed as:
\begin{equation}
\label{E:tausol}
 \tau{(t,x^i)}
=\int\frac{d^pk}{(2\pi)^p}\left[\tau_1(k_i)h^{(1)}(t,k)
e^{ik_ix^i}+\tau_2(k_i)h^{(2)}(t,k)e^{-ik_ix^i}\right],
\end{equation}
where $\tau_1(k_i)$, $\tau_2(k_i)$ are constant parameters
corresponding to the mode $k_i$ and
\begin{equation}
 h^{(1,2)}(t,k)=e^{-\beta t}H^{(1,2)}_1(2\ka e^{-\beta t}).
\end{equation}
$H^{(1,2)}_1(s)$ are the Hankel function, which are the linear
combinations of the Bessel function of the first kind $J_1(s)$ and
the second kind $Y_1(s)$: $H^{(1)}_1(s)=J_1(s)+iY_1(s)$ and
$H^{(2)}_1(s)=J_1(s)-iY_1(s)$. They satisfy
$H^{(1)}_1(s)=H^{(2)}_1(s)^*$. In order to get $\tau=\tau^*$,
$\tau_1(k_i)=\tau_2(k_i)^*$. Then we can set
$\tau_1(k_i)=\tau_0(k_i)e^{i\delta(k_i)}$ and
$\tau_2(k_i)=\tau_0(k_i)e^{-i\delta(k_i)}$.

Here, we consider the real tachyon field $T$. The perturbation
(\ref{E:tausol}) should be real as well and can be expressed as,
denoting $s=2\ka e^{-\beta t}$:
\begin{equation}\label{E:persol}
 \tau{(t,x^i)}=\int \frac{d^pk}{(2\pi)^p}\frac{s}{2\ka}
\tau_0(k_i)\left[J_1(s)\cos{(k_ix^i+\delta(k_i))
-Y_1(s)\sin{(k_ix^i+\delta(k_i))}}\right].
\end{equation}

For small parameters $0<s\ll \sqrt{2}$,
\begin{eqnarray}
 J_1(s)\simeq \frac{s}{4}, \textrm{ }\textrm{ }
\textrm{ } Y_1(s)\simeq -\frac{2}{\pi s}.
\end{eqnarray}
Thus, at late time, Eq.\ (\ref{E:persol}) approximately reduces
to:
\begin{equation}
 \tau\simeq\frac{\beta}{\pi}\int\frac{d^pk}
{(2\pi)^p}\frac{\tau_0(k_i)}{k}\sin{(k_ix^i+\delta(k_i))}.
\end{equation}
The result indicates that the perturbation corresponding to each
mode $k_i$ freezes at late time and that shorter wavelength modes
are damped more than longer wavelength.

\subsection{Comparison with the numerical simulations}

We make numerical simulations of the perturbations in $1+1$
dimensions based on the Hamiltonian formalism
\cite{Gibbons:2000hf,Gibbons:2002tv,Sen:2002an}. Define the
momentum conjugate to $T$
\begin{equation}
 \Pi=\frac{\delta S}{\delta\dot{T}}=\frac{V(T)\dot{T}}
{\sqrt{1+y}}.
\end{equation}
Then the equations of motion can be expressed as
\begin{equation}\label{e:hamdotPi}
 \dot{\Pi}=\nabla^j\left(\frac{\nabla_jT\sqrt{\Pi^2+V^2}}
{\sqrt{1+(\nabla T)^2}}\right)-\frac{VV'\sqrt{1+
(\nabla T)^2}}{\sqrt{\Pi^2+V^2}},
\end{equation}
\begin{equation}\label{e:hamdotT}
 \dot{T}=\frac{\Pi\sqrt{1+(\nabla T)^2}}{\sqrt{\Pi^2+V^2}}.
\end{equation}

The simulation is implemented on a spatial lattice of $1280$
points. The lattice spacing and the timestep are set to be
respectively: $\delta x=0.05$ and $\delta t=0.01$. We use the
symmetric difference for the first order field derivatives and a
3-point stencil for the second order field derivatives. We adopt
the second order Runge-Kutta method for the time update. In the
simulation, the constant $\beta=1/\sqrt{2}$.

To get the perturbation, we first need to produce two sets of
simulation data with two different initial conditions
respectively: $T(t=0,x)=D$ and $T(t=0,x)=D+\delta(x)$, where $D$
is the field value at $t=0$: $D=T_0(t=0)$, and $\delta(x)$ is the
random initial perturbation: $\delta(x)=\tau(t=0,x)$. The
difference between them gives the $1+1$ dimensional evolution
surface of $\tau(t,x)$. In Fig.\ \ref{f:tausim}, we present the
plots of the perturbation at $t=8,24,38.4$ respectively. We can
see that the perturbation almost ``freezes'', especially at peaks
and troughs, which is consistent with the linear perturbation
analysis in the previous subsection. However, the second
derivative $\nabla^2\tau$ can be seen to decrease in magnitude
near peaks, and increase near troughs, until the simulation breaks
down (signalled by a spike in $\tau$ at around $x\simeq27$ and
$x\simeq53$). These features are consistent with the analysis in
Section 3, but not with the predictions from the linear
perturbation equation. Hence, we need to take into account higher
order perturbation terms to understand this behaviour.

\FIGURE[ht]{
\epsfig{file=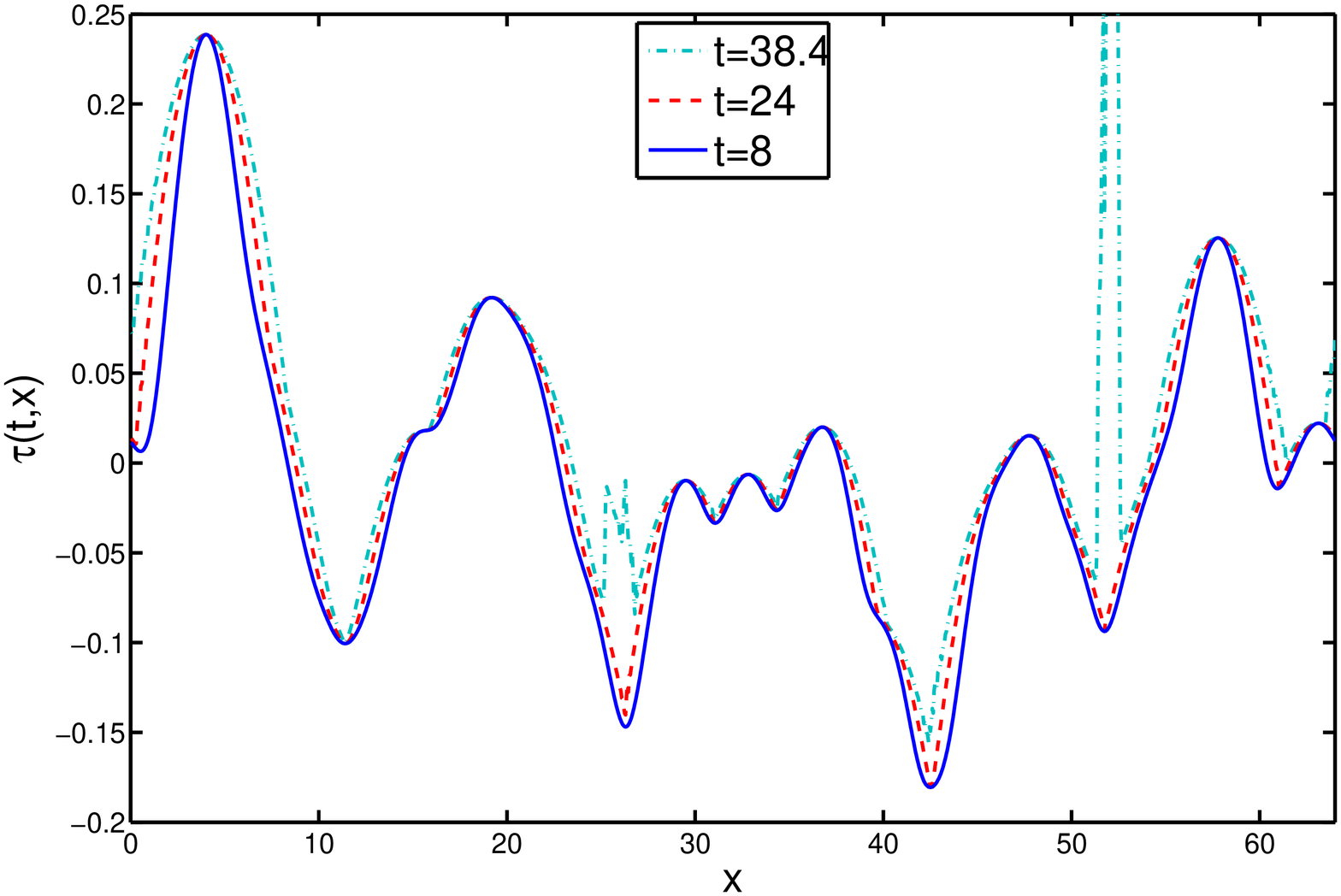,scale=0.4}
\caption{The numerical plots of the perturbation $\tau(t,x)$
based on Eqs.\ (\ref{e:hamdotPi}) and (\ref{e:hamdotT}). The
initial homogeneous field is set to be $D=1$.
\label{f:tausim}}}


\subsection{Non-linear perturbations and development of caustics}

We start with a discussion of the behaviour of $1+y$. For large
$T$, we can assume that $\dot{T}\rightarrow 1$,
$\ddot{T}\rightarrow 0$ and $V'/V\simeq -\beta$. The equation of
motion (\ref{e:feq1}) becomes
\begin{equation}\label{E:appEoM}
 \frac{\partial}{\partial t}(1+y)\simeq -2(\nabla^2 T
+\beta)(1+y)+2\nabla_i T\nabla_j T\nabla_i\nabla_j T
-\dot{T}\frac{\partial}{\partial t}(\nabla^2 T).
\end{equation}
If the last two terms are negligible compared to the first one,
then $1+y$ should decrease exponentially with time to zero. This
accounts for the numerical observation first made in
\cite{Felder:2002sv}.

To check the expectation that $1+y\simeq0$, we present the plots
of $1+y$ in Fig.\ (\ref{f:persim}) for a field
$T(t,x)=D+t+\tau(t,x)$, where the plots of the perturbation $\tau$
are given in Fig.\ (\ref{f:tausim}). The result indicates that
$1+y$ really decreases with time to very small positive values
exponentially in the early stage. However, the simulation can not
continue at late time due to instabilities emerging at the
positions where $\tau$ eventually gets discontinous in Fig.\
(\ref{f:tausim}), i.e., near troughs. This implies that the
instability should correlate with the caustics formation. In what
follows, we will determine how it happens in the perturbation
method .

\FIGURE[ht]{
\epsfig{file=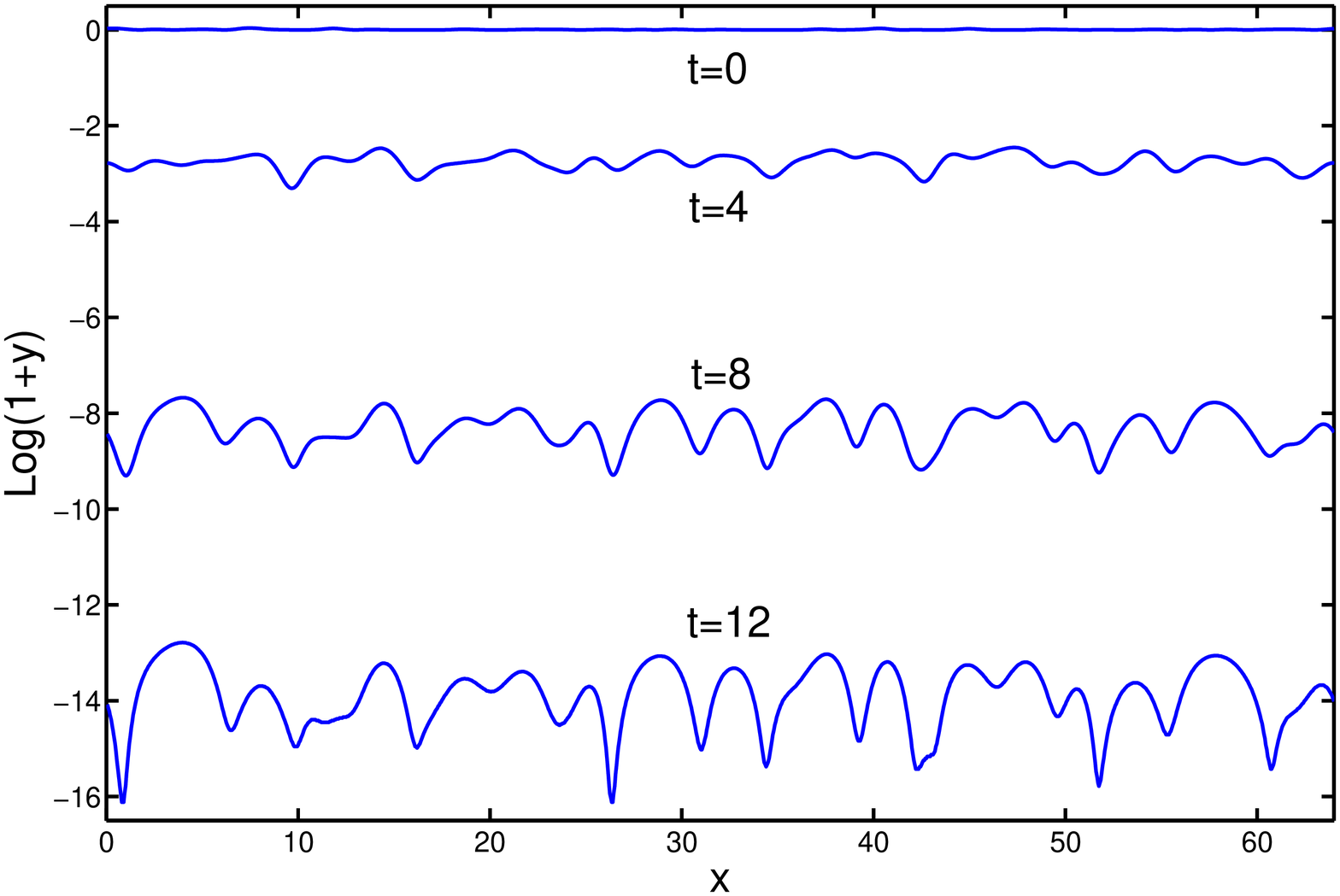,scale=0.4}
\caption{ The plots of $\log(1+y)$ for the field
$T(t,x)=D+t+\tau(t,x)$, where $D=1$ and the plots of $\tau(t,x)$
are given in Fig.\ (\ref{f:tausim}). \label{f:persim}} }


Including the quadratic order terms of $\tau$ in the second square
bracket in Eq.\ (\ref{E:pereq}), we have in the limit
$\dot{T}_0\rightarrow1$
\begin{equation}\label{e:quadperteq}
 \partial_t[2\dot{\tau}-(\nabla\tau)^2]=
-\be[2\dot{\tau}-(\nabla\tau)^2]+\ddot{\tau}
-\dot{\tau}(2\nabla^2\tau+\be\dot{\tau}).
\end{equation}
It is consistent with Eq.\ (\ref{E:bt1}) if setting
$\tau=b(t)\triangle x^2/2$ in $p=1$ case. When the last two terms
on the right hand side of the equation are less important than the
first one, we will get at late time
\begin{equation}
 1+y\simeq 2\dot{\tau}-(\nabla\tau)^2=0.
\end{equation}
Specially, $\dot{\tau}=0$ when $\nabla_i\tau=0$, i.e., at exactly
the peaks or troughs, the perturbation ``freezes''. But away the
peaks and troughs, the perturbation increases with a increasing
rate equal to $(\nabla\tau)^2/2$.

Let us see what features we can have from the equation
$\dot{\tau}=(\nabla\tau)^2/2$ around peaks and troughs
respectively. Since $\tau$ ``freezes'' at exactly the peaks and
troughs and grows away from them, the field gradient
$\nabla_i\tau$ decreases near peaks and increases near troughs.
Therefore, the curves of $\tau$ tend to flatten near peaks untill
$\nabla_i\tau=0$, where $\dot{\tau}=0$ and so $\tau$ stops
growing. While near troughs, the curves of $\tau$ sharpen, with
$|\nabla_i\tau|$ increasing. This accelerates the growth of
$\tau$. In this case, $\tau$, $\nabla_i\tau$ and $\nabla^2\tau$
all will tend to infinity, which accounts for the caustic
formation and so the instability near the vacuum.

\section{The BSFT action}
\label{sec:bsftact}

The effective action derived from BSFT is given by
\cite{Kutasov:2000aq,Kraus:2000nj}:
$\mathcal{L}=-V(T)F(y)$ with
the potential
\begin{equation}
\label{E:Pot2}
 V(T)=\exp{\left(-\frac{\beta^2 T^2}{2}\right)},
\end{equation}
and the kinetic part
\begin{equation}\label{e:bsftact}
 F(y)=\frac{1}{2}\frac{4^yy\Gamma^2(y)}
{\Gamma(2y)}=2^{2y-1}yB(y),
\end{equation}
where $y=\partial^\mu T\partial_\mu T$ as before and $B(y)$ is the
beta function. It is easy to see that $F\geq 0$ when $y\geq-1/2$.
Aspects of the dynamics arising from the the BSFT action have been
given in \cite{Sugimoto:2002fp,Barnaby:2004nk}.

Using $\Gamma(1+u)=u\Gamma(u)$, we can derive the recursion
relation
\begin{eqnarray}\label{e:betarec}
 B(y)=2^{2m}\frac{(y+\frac{2m-1}{2})(y+\frac{2m-3}{2})\cdots(y
+\frac{1}{2})}{(y+m-1)(y+m-2)\cdots(y+1)y}B(y+m), & m=1,2,3,\cdots
\end{eqnarray}
From Stirling's approximation, $B(u)\simeq 2^{1-2u}\sqrt{\pi/u}$
for large $u$. Then the BSFT action can be expressed as
\begin{equation}\label{e:appbsftact}
 F(y)\simeq \frac{(y+\frac{2m-1}{2})(y+\frac{2m-3}{2})\cdots(y+
\frac{1}{2})}{(y+m-1)(y+m-2)\cdots(y+1)}\sqrt{\frac{\pi}{y+m}},
\end{equation}
and we will find it useful to take $m\gg-y$.

For $m=1$, we approximately have
\begin{equation}
 F(y)\simeq\sqrt{\pi}\sqrt{1+y},
\end{equation}
when $1+y\gg 1$. This is the DBI action. So the behaviour of the
BSFT action (\ref{e:bsftact}) for large $1+y$ should be similar to
that discussed in previous sections. In what follows, we will
discuss the behaviour near the vacuum when $y$ is small or
negative.

\subsection{Time evolution}

We first explore the homogeneous approach to the vacuum. The
equation of motion in this case is
\begin{equation}\label{e:bsfthomeq}
 F-2yF'=\frac{E}{V},
\end{equation}
where $y=-\dot{T}^2$, $F'=\partial F/\partial y$ and $E$ is a
constant.

The condensation starts with a small velocity $|\dot{T}|$. Let us
consider the case when $\dot{T}=0$. With the relation of the gamma
function $\Gamma(2u)=2^{2u-1}\Gamma(u)\Gamma(u+1/2)/\sqrt{\pi}$,
we can rewrite the kinetic part $F(y)$ as
\begin{equation}
 F(y)=\sqrt{\pi}\frac{\Gamma(y+1)}{\Gamma(y+\frac{1}{2})}.
\end{equation}
Thus, $F(0)=1$ since $\Gamma(1/2)=\sqrt{\pi}$. This gives $V(T)=E$
at $\dot{T}=0$ in terms of the equation of motion.

Now we need to find where the tachyon potential gets the minimum
value $V(T)=0$, which corresponds to the vacuum. Since
$y=-\dot{T}^2\leq 0$ in the homogeneous case, we can change the
action to the form by using $\Gamma(1-u)\Gamma(u)=\pi/\sin(\pi u)$
\begin{equation}
 F(y)=-2^{2y+1}\pi\cot(\pi y)\frac{1}{B(-y)}\simeq-\sqrt{\pi}
\cot(\pi y)\sqrt{-y}.
\end{equation}
Then the homogeneous equation of motion (\ref{e:bsfthomeq})
becomes
\begin{equation}\label{e:appbsfthomeq}
 \frac{2(-\pi y)^{\frac{3}{2}}}{\sin^2(\pi y)}\simeq\frac{E}{V}.
\end{equation}
It is easily seen that $V=0$ at $y=-n$, where $n=1,2,3,\cdots$.
The behaviour in taking the limit $\dot{T}^2\rightarrow n$ is,
\begin{equation}
 V(T)\sim\frac{E}{2}\sqrt{\frac{\pi}{n^3}}(\dot{T}^2-n)^2.
\end{equation}
The equation implies that the tachyon potential vanishes for
either $\dot{T}^2\rightarrow n$ or $n\rightarrow\infty$. We will
assume that only the first pole ($y=-1$) in Eq.\
(\ref{e:appbsfthomeq}) is relevant, as the condensation starts
with $y\simeq 0$.

\subsection{Dynamics near the vacuum}

The full equation of motion for a general $F(y)$ is
\begin{equation}
 \left(2\frac{F''}{F'}\partial_\mu T\partial_\nu T+
\eta_{\mu\nu}\right)\partial^\mu\partial^\nu T+
\frac{V'}{V}\left(y-\frac{1}{2} \frac{F}{F'}\right)=0.
\end{equation}
Note that $V'=\partial V/\partial T$, $F'=\partial F/\partial y$
and $F''=\partial^2 F/\partial y^2$.

We now consider the dynamics approaching the vacuum when
$y\rightarrow -1$ from above.  Since $F/F'=1/(\ln F)'$ and
$F''/F'=(\ln F)''/(\ln F)'+(\ln F)'$, we only need to know $(\ln
F)'$ and $(\ln F)''$ for deriving the equation of motion. In the
limit $y\rightarrow-1$, we approximately have for any $m\geq 2$
from Eq.\ (\ref{e:appbsftact})
\begin{equation}
 (\ln F)'\sim-\frac{1}{1+y}, \textrm{ }\textrm{ }
\textrm{ } (\ln F)''\sim\frac{1}{(1+y)^2}.
\end{equation}
The corresponding expressions for the DBI action are
$(\ln F)'=1/[2(1+y)]$ and $(\ln F)''=-1/[2(1+y)^2]$.
Inserting them in the equation of motion, we have
\begin{equation}
 \left(\Box T-\frac{V'}{V}\right)(1+y)
\simeq 2\partial^\mu T\partial_\mu(1+y),
\end{equation}
So it is very similar to Eq.\ (\ref{e:feq1}) for the DBI action.
The only difference is a change of the coefficient on the right
hand side. Therefore, there should be similar consequences for the
dynamics approaching the vacuum from the BSFT action to those from
the DBI action. That is, the perturbations ``freeze'' and the
second derivative becomes discontinuous.

\section{Tachyon matter}
\label{sec:tm}

Reverting to the DBI action, the energy density and the pressure
are given in flat spacetime by Eq.\ (\ref{E:enemom}):
\begin{equation}
 \rho=T_{00}=\frac{1+(\nabla T)^2}{\sqrt{1+y}}V(T),
\end{equation}
\begin{equation}\label{e:pi}
 p_i=T_{ii}=-V(T)\frac{1+y-(\nabla_iT)^2}{\sqrt{1+y}}=
q+\frac{(\nabla_iT)^2}{1+(\nabla T)^2}\rho,
\end{equation}
where there is no sum on $i$ and $q=\mathcal{L}=-V(T)\sqrt{1+y}$,
and so
\begin{equation}
 w_i=\frac{p_i}{\rho}=-\frac{1+y}{1+(\nabla T)^2}+
\frac{(\nabla_iT)^2}{1+(\nabla T)^2}.
\end{equation}

From the above formulas, it is easy to see that the system at the
beginning of the tachyon condensation has positive energy density
and negative pressure everywhere: $p_i=-\rho=-V_m$, the
characteristic of the interior of a brane. After the condensation
starts, the behavior in different areas becomes different.

Consider the case that the decay of the D$p$-brane happens only in
the $x^1$ direction to produce a D$(p-1)$-brane. Then the decay
process can be described by a kink-anti-kink tachyon solution
along the $x^1$ direction.

First, we consider the appearance of a D$(p-1)$-brane, which we
locally choose to be orthogonal to the $x^1$-direction. Since
$\dot{T}=0$ and $|\nabla_1T|\rightarrow\infty$ towards the end of
the decay at kinks, the energy density near the kinks
$\rho_K\rightarrow\infty$. If the approximate solution found in
Sec. 3.1 is to be believed, this divergence happens in finite
time, in accordance with the analysis in the string field theory
\cite{Sen:2002nu,Larsen:2002wc}. The pressure components in all
$p$ spatial directions are respectively: $p_{jK}=q$ ($2\leq j\leq
p$) and
\begin{equation}
 p_{1K}=p_{jK}+\frac{(\nabla_1T)^2}{1+(\nabla_1T)^2}\rho_K.
\end{equation}
Therefore, $p_{jK}\simeq-\rho_K$ and $p_{1K}\simeq p_{jK}+\rho_K$
near the end of the condensation. And so
$p_{jK}\rightarrow-\infty$ and $p_{1K}\rightarrow 0$ as
$t\rightarrow\infty$, which further give $w_{jK}\rightarrow -1$
and $w_{1K}\rightarrow 0$. This is consistent with the appearance
of a D$(p-1)$-brane in a vacuum.

Second, we consider the approach to vacuum of a homogeneous field
with an expanded inhomogeneous part. The energy density $\rho_V$
and the pressure $p_{iV}$ near the vacuum can be analysed using
the results in Section 3. The final values of $\rho_V$ and
$p_{iV}$ at the end of the condensation are not obviously
determined because the two quantities $V(T)$ and $1+y$ both tend
to zero as $t\rightarrow\infty$. They can be estimated in terms of
the expression of $1+y$ in Eq.\ (\ref{E:Rel}), from which we see
that $1+y\simeq Ae^{-2(\be+b)t}\rightarrow Ae^{-2\be t}$ when
$b<0$ (e.g., near peaks) and $1+y\rightarrow-C\triangle
x^2e^{-2\be t}$ when $b>0$ (e.g., near troughs). To keep $1+y$
positive, we assume that $C=\dot{b}_0-b_0^2$ is negative. For the
inverse hyperbolic potential (\ref{E:Pot1}), $V(T)\simeq
2V_m\exp{[-\beta(D+t)]}$. Thus the energy density near the vacuum
is
\begin{equation}
 \rho_V\simeq
 \left\{
 \begin{array}{cl}
 \frac{2V_m}{\sqrt{A}}e^{-\be D}, & (b<0), \\
 \frac{2V_m(1+b^2\triangle x^2)}{\sqrt{Ae^{-2bt}
-C\triangle x^2}}e^{-\beta D}, & (b>0).
 \end{array}
 \right.
\end{equation}
Thus the energy density tends to constant near peaks and diverges
at troughs $\triangle x=0$ as $t\rightarrow\infty$. When $b$ is
not too large, $\rho_V\sim1/|\triangle x|$ near but away from
the trough at $\triangle x=0$.

In Fig.\ \ref{f:energy}, we present the plots of the energy
density $\rho_V$ and its inverse $1/\rho_V$ relative to the plots
of the second derivative $b=T''$ at a time $t=16$. The figure
indicates strong relation between $\rho_V$ and $T''$: the energy
density peaks where $T''$ also peaks. By comparison with Fig.\
\ref{f:tausim}, we see that the sharpest peaks in $T''$ occur at
troughs in $T$. The top plot shows evidence that around the peaks
of the energy density, we have the approximate relation
$1/\rho_V\sim|\triangle x|$.

\FIGURE[ht]{
\epsfig{file=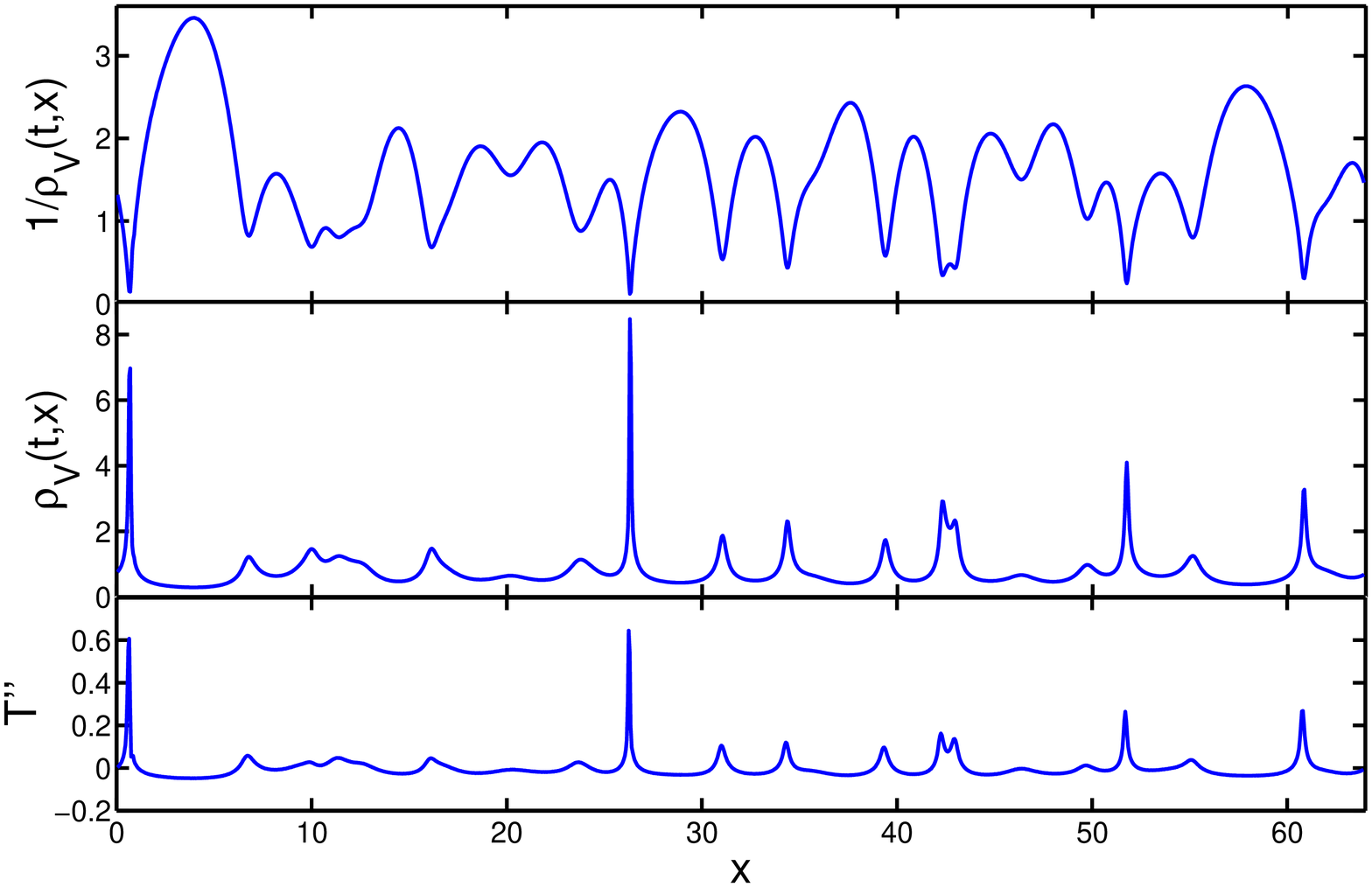,scale=0.4}
\caption{The correlation between the energy density $\rho_V(t,x)$
near the vacuum and the second derivative $T''$. The plots are made
at $t=16$. It is clear to see that $\rho_V(t,x)$ peaks at positions
where $T''$ also peaks and that $1/\rho_V\sim|\triangle x|$ near
the peaks of $\rho_V$. \label{f:energy}} }


It is interesting to note that the tachyon fluid is not quite
pressureless in general. Indeed, from Eq.\ (\ref{e:pi}), noting
that $q\rightarrow 0$ near the vacuum as $t\rightarrow\infty$, we
find $\rho_V$: $p_{iV}\sim\rho_V(\nabla_iT)^2/[1+(\nabla T)^2]$.
Given that $T$ ``freezes", we see that the small fluctuations give
rise to a small pressure. The equation of state parameter $w_{iV}$
is non-zero.

\section{The generalised DBI action}
\label{sec:genDBI}

The problem with the divergence in the gradient of the field near
kinks has been recognised for some time
\cite{Felder:2002sv,Cline:2003vc,Barnaby:2004dz}, and
regularisations of the action proposed
\cite{Brax:2003rs,Copeland:2003df} with respect to this problem,
which have the important property that their static kinks have
finite field gradients as the minimum potential $V_0\rightarrow
0$. In what follows, we will consider the modified action given in
\cite{Brax:2003rs} in the $1+1$ dimensions:
\begin{equation}
 \mathcal{L}=-V(T)(1+y)^{\frac{1}{2}(1+\epsilon)}.
\end{equation}
where $\epsilon$ is a small positive value. The equation of motion
from it is:
\begin{eqnarray}
\label{E:GEoM}
 \ddot{T} & = & f\{2(1-\epsilon^2)\dot{T}\nabla_i{T}
\nabla^i{\dot{T}}-\frac{V'}{V}(1+y)(1-\epsilon y) \nonumber \\
 & &+(1+\epsilon)[(1+y)\nabla^2T-(1-\epsilon)\nabla_i{T}
\nabla_j{T}\nabla^i\nabla^j{T}]\},
\end{eqnarray}
where $f=1/\left[(1+\epsilon)(1+\nabla_iT\nabla^iT-\epsilon
\dot{T}^2)\right]$.

First, We consider the field expansion around the kinks and
antikinks in the case with the hyperbolic potential
(\ref{E:Pot1}). Doing the field expansion like (\ref{E:Exp1}), we
get the equation of $a(t)$:
\begin{equation}
\label{E:GaE}
 \ddot{a}=\frac{\beta^2 a}{1+\epsilon}-\frac{\epsilon\beta^2a^3}
{1+\epsilon} +\frac{(1-\epsilon)(2a\dot{a}^2+c)}{1+a^2}+\epsilon c.
\end{equation}
So when $a$ grows large and satisfies $1\ll a\ll
1/\sqrt{\epsilon}$, we have a similar solution to Eq.\
(\ref{E:aSol}). When $a\gg 1/\sqrt{\epsilon}$, the $a^3$ term
becomes important and $a$ begins to execute damped oscillations.
$a$ settles to $a=1/\sqrt{\ep}$.

To show this point clearly, we can transform Eq.\ (\ref{E:GaE})
into a soluble form, assuming $1+a^2\simeq a^2$, and neglecting
$c$. When $\epsilon \neq 1/2$, we set $a=z^{-1/(1-2\epsilon)}$ and
then have:
\begin{equation}
\label{E:Gzt}
 \ddot{z}+\frac{\partial{U(z)}}{\partial{z}}=0,
\end{equation}
where $U(z)$ is:
\begin{equation}
 U(z)=\frac{\left(\frac{1}{2}-\epsilon\right)\beta^2z^2}
{(1+\epsilon)}\left[1+\left(\frac{1}{2}-\epsilon\right)
z^{-\frac{2}{1-2\epsilon}}\right],
\end{equation}
For $\epsilon > 0$, the minimum value of $U(z)$ happens at
$z=\epsilon^{(1/2-\epsilon)}$ or
\begin{equation}
 a=\frac{1}{\sqrt{\epsilon}}.
\end{equation}
This value should be the final field gradient
$T'\simeq a$ as $t\rightarrow \infty$. When $\epsilon=1/2$,
we can set $a=e^z$ and get the same equation
as in (\ref{E:Gzt}) but with a different potential $U(z)$:
\begin{equation}
 U(z)=\frac{\beta^2}{6}\left(e^{2z}-4z\right).
\end{equation}
The position according for the minimum potential $U(z)$ is
$z=\ln\sqrt{2}$ or $a=\sqrt{2}$, which is consistent to the above
result of $\epsilon\neq 1/2$.

For the expansion around extrema, we get:
\begin{equation}
 \ddot{T}_0=\left[\frac{1+\epsilon\dot{T}_0^2}{1+\epsilon}\beta
+b\right]\frac{1-\dot{T}_0^2}{1-\epsilon\dot{T}_0^2}.
\end{equation}
When $\epsilon<1$, $1-\epsilon\dot{T}_0^2$ is always positive if
$|\dot{T}_0|\leq 1$. Similar to the analysis for the DBI action
with $\epsilon=0$, $|\dot{T}_0|$ still goes to $1$ here at the end
of the condensation.

With the approximation $\dot{T}_0\simeq 1$ for positive $T_0$, the
equation for $b(t)$ can be expressed as:
\begin{equation}
 (1-\epsilon)\partial_t{(\dot{b}-b^2)}=-2(\beta-\epsilon b)
(\dot{b}-b^2).
\end{equation}
For $\epsilon<1$, it can be shown both analytically and
numerically that there are still two possible final values for
$b(t)$ at $t\rightarrow\infty$. When $b$ tends to zero, we can
write the solution similar to (\ref{E:bt1}). So it can evolve to
zero at $t\rightarrow\infty$. When
$b(t\rightarrow\infty)\rightarrow\infty$, $b$ grows to infinity
more quickly than the case with $\epsilon=0$. Thus caustics are
possible to form in a finite time for the modified DBI action.

Hence, although this modified effective action solves the problem
of diverging gradients near kinks, it still has unstable solutions
and breaks down as the vacuum is approached.

\section{Conclusions}
\label{sec:conclusions}

We have investigated the inhomogeneous tachyon condensation
process, focusing on the solutions near kinks (where $T=0$) and
also in regions where the field is approximately uniform.

We obtained an approximate space-time dependent solution near the
kinks and anti-kinks, verifying that the spatial derivative of the
field $T'$ tends to infinity in a finite time. We show that this
singularity can be avoided by modifying the effective action in
$1+1$ dimensions, which is consistent with the result obtained in
the static case by \cite{Brax:2003rs,Copeland:2003df}.

We then studied inhomogeneous tachyon condensation in the absence
of kinks. In (\ref{E:Rel}), an analysis assuming that the tachyon
field obeyed the eikonal equation $\dot{T}^2+\nabla T^2=1$
showed that near troughs $T''$ diverges in finite time, which was
interpreted as the production of caustics in free-streaming
matter. Directly from the semi-analytical solutions, we learnt how
the field obeys this equation and how the solutions from this
equation lead to singularities in $T''$. A linear analysis shows
that perturbations ``freeze'', except near extrema; Adding in
higher order terms shows that the curvative ($\nabla^2T$)
increases near troughs (extrema with $\nabla^2T>0$) and decreases
near peaks ($\nabla^2T<0$). For  the former case, this leads to
discontinious gradients near troughs, which in the eikonal
approximation leads to caustics. Moreover, the energy density
$\rho_V$ diverges as $t\rightarrow\infty$ near troughs. The
pressure of tachyon matter is small, but the equation of state
parameter $w$ does not generically vanish.

The same analysis applies also to BSFT effective theories, and to
another simple proposal for modifying the effective action
\cite{Brax:2003rs,Copeland:2003df}. The large gradients and higher
derivatives signal a breakdown of the effective action
(\ref{E:DBIA}) for describing the dynamics of the tachyon field
near the vacuum. Whether or not caustics actually form remains
open for further discussion.

\acknowledgments
HL is supported by a Dorothy Hodgkin Postgraduate Award.

\newpage
\bibliographystyle{JHEP}
\bibliography{b}

\providecommand{\href}[2]{#2}\begingroup\raggedright\begin{thebibliography}{10}

\bibitem{Garousi:2000tr}
M.~R. Garousi, {\it Tachyon couplings on non-bps d-branes and dirac-born-infeld
  action},  {\em Nucl. Phys.} {\bf B584} (2000) 284--299,
  [\href{http://xxx.lanl.gov/abs/hep-th/0003122}{{\tt hep-th/0003122}}].

\bibitem{Bergshoeff:2000dq}
E.~A. Bergshoeff, M.~de~Roo, T.~C. de~Wit, E.~Eyras, and S.~Panda, {\it
  T-duality and actions for non-bps d-branes},  {\em JHEP} {\bf 05} (2000) 009,
  [\href{http://xxx.lanl.gov/abs/hep-th/0003221}{{\tt hep-th/0003221}}].

\bibitem{Sen:2002an}
A.~Sen, {\it Field theory of tachyon matter},  {\em Mod. Phys. Lett.} {\bf A17}
  (2002) 1797--1804, [\href{http://xxx.lanl.gov/abs/hep-th/0204143}{{\tt
  hep-th/0204143}}].

\bibitem{Hashimoto:2002ct}
K.~Hashimoto and N.~Sakai, {\it Brane - antibrane as a defect of tachyon
  condensation},  {\em JHEP} {\bf 12} (2002) 064,
  [\href{http://xxx.lanl.gov/abs/hep-th/0209232}{{\tt hep-th/0209232}}].

\bibitem{Brax:2003rs}
P.~Brax, J.~Mourad, and D.~A. Steer, {\it Tachyon kinks on non bps d-branes},
  {\em Phys. Lett.} {\bf B575} (2003) 115--125,
  [\href{http://xxx.lanl.gov/abs/hep-th/0304197}{{\tt hep-th/0304197}}].

\bibitem{Kim:2003in}
C.-j. Kim, Y.-b. Kim, and C.~O. Lee, {\it Tachyon kinks},  {\em JHEP} {\bf 05}
  (2003) 020, [\href{http://xxx.lanl.gov/abs/hep-th/0304180}{{\tt
  hep-th/0304180}}].

\bibitem{Hindmarsh:2007dz}
M.~Hindmarsh and H.~Li, {\it {Perturbations and moduli space dynamics of
  tachyon kinks}},  {\em Phys. Rev.} {\bf D77} (2008) 066005,
  [\href{http://xxx.lanl.gov/abs/0711.0678}{{\tt arXiv:0711.0678}}].

\bibitem{Cline:2003vc}
J.~M. Cline and H.~Firouzjahi, {\it Real-time d-brane condensation},  {\em
  Phys. Lett.} {\bf B564} (2003) 255--260,
  [\href{http://xxx.lanl.gov/abs/hep-th/0301101}{{\tt hep-th/0301101}}].

\bibitem{Felder:2002sv}
G.~N. Felder, L.~Kofman, and A.~Starobinsky, {\it Caustics in tachyon matter
  and other born-infeld scalars},  {\em JHEP} {\bf 09} (2002) 026,
  [\href{http://xxx.lanl.gov/abs/hep-th/0208019}{{\tt hep-th/0208019}}].

\bibitem{Barnaby:2004dz}
N.~Barnaby, A.~Berndsen, J.~M. Cline, and H.~Stoica, {\it Overproduction of
  cosmic superstrings},  {\em JHEP} {\bf 06} (2005) 075,
  [\href{http://xxx.lanl.gov/abs/hep-th/0412095}{{\tt hep-th/0412095}}].

\bibitem{Mazumdar:2001mm}
A.~Mazumdar, S.~Panda, and A.~Perez-Lorenzana, {\it Assisted inflation via
  tachyon condensation},  {\em Nucl. Phys.} {\bf B614} (2001) 101--116,
  [\href{http://xxx.lanl.gov/abs/hep-ph/0107058}{{\tt hep-ph/0107058}}].

\bibitem{Fairbairn:2002yp}
M.~Fairbairn and M.~H.~G. Tytgat, {\it Inflation from a tachyon fluid?},  {\em
  Phys. Lett.} {\bf B546} (2002) 1--7,
  [\href{http://xxx.lanl.gov/abs/hep-th/0204070}{{\tt hep-th/0204070}}].

\bibitem{Choudhury:2002xu}
D.~Choudhury, D.~Ghoshal, D.~P. Jatkar, and S.~Panda, {\it On the cosmological
  relevance of the tachyon},  {\em Phys. Lett.} {\bf B544} (2002) 231--238,
  [\href{http://xxx.lanl.gov/abs/hep-th/0204204}{{\tt hep-th/0204204}}].

\bibitem{Kofman:2002rh}
L.~Kofman and A.~Linde, {\it Problems with tachyon inflation},  {\em JHEP} {\bf
  07} (2002) 004, [\href{http://xxx.lanl.gov/abs/hep-th/0205121}{{\tt
  hep-th/0205121}}].

\bibitem{Sami:2002fs}
M.~Sami, P.~Chingangbam, and T.~Qureshi, {\it Aspects of tachyonic inflation
  with exponential potential},  {\em Phys. Rev.} {\bf D66} (2002) 043530,
  [\href{http://xxx.lanl.gov/abs/hep-th/0205179}{{\tt hep-th/0205179}}].

\bibitem{Shiu:2002qe}
G.~Shiu and I.~Wasserman, {\it Cosmological constraints on tachyon matter},
  {\em Phys. Lett.} {\bf B541} (2002) 6--15,
  [\href{http://xxx.lanl.gov/abs/hep-th/0205003}{{\tt hep-th/0205003}}].

\bibitem{Steer:2003yu}
D.~A. Steer and F.~Vernizzi, {\it {Tachyon inflation: Tests and comparison with
  single scalar field inflation}},  {\em Phys. Rev.} {\bf D70} (2004) 043527,
  [\href{http://xxx.lanl.gov/abs/hep-th/0310139}{{\tt hep-th/0310139}}].

\bibitem{Dvali:1998pa}
G.~R. Dvali and S.~H.~H. Tye, {\it Brane inflation},  {\em Phys. Lett.} {\bf
  B450} (1999) 72--82, [\href{http://xxx.lanl.gov/abs/hep-ph/9812483}{{\tt
  hep-ph/9812483}}].

\bibitem{Sarangi:2002yt}
S.~Sarangi and S.~H.~H. Tye, {\it Cosmic string production towards the end of
  brane inflation},  {\em Phys. Lett.} {\bf B536} (2002) 185--192,
  [\href{http://xxx.lanl.gov/abs/hep-th/0204074}{{\tt hep-th/0204074}}].

\bibitem{Choudhury:2003vr}
D.~Choudhury, D.~Ghoshal, D.~P. Jatkar, and S.~Panda, {\it Hybrid inflation and
  brane-antibrane system},  {\em JCAP} {\bf 0307} (2003) 009,
  [\href{http://xxx.lanl.gov/abs/hep-th/0305104}{{\tt hep-th/0305104}}].

\bibitem{Sen:2002in}
A.~Sen, {\it Tachyon matter},  {\em JHEP} {\bf 07} (2002) 065,
  [\href{http://xxx.lanl.gov/abs/hep-th/0203265}{{\tt hep-th/0203265}}].

\bibitem{Frolov:2002rr}
A.~V. Frolov, L.~Kofman, and A.~A. Starobinsky, {\it {Prospects and problems of
  tachyon matter cosmology}},  {\em Phys. Lett.} {\bf B545} (2002) 8--16,
  [\href{http://xxx.lanl.gov/abs/hep-th/0204187}{{\tt hep-th/0204187}}].

\bibitem{Kutasov:2000aq}
D.~Kutasov, M.~Marino, and G.~W. Moore, {\it {Remarks on tachyon condensation
  in superstring field theory}},
  \href{http://xxx.lanl.gov/abs/hep-th/0010108}{{\tt hep-th/0010108}}.

\bibitem{Kraus:2000nj}
P.~Kraus and F.~Larsen, {\it {Boundary string field theory of the DD-bar
  system}},  {\em Phys. Rev.} {\bf D63} (2001) 106004,
  [\href{http://xxx.lanl.gov/abs/hep-th/0012198}{{\tt hep-th/0012198}}].

\bibitem{Kutasov:2003er}
D.~Kutasov and V.~Niarchos, {\it Tachyon effective actions in open string
  theory},  {\em Nucl. Phys.} {\bf B666} (2003) 56--70,
  [\href{http://xxx.lanl.gov/abs/hep-th/0304045}{{\tt hep-th/0304045}}].

\bibitem{Smedback:2003ur}
M.~Smedback, {\it {On effective actions for the bosonic tachyon}},  {\em JHEP}
  {\bf 11} (2003) 067, [\href{http://xxx.lanl.gov/abs/hep-th/0310138}{{\tt
  hep-th/0310138}}].

\bibitem{Jeffrey:2007}
A.~Jeffrey and D.~Zwillinger, {\em Table of Integrals, Series, and Products}.
\newblock 2007.
\newblock Academic Press.

\bibitem{abramowitz+stegun}
M.~Abramowitz and I.~A. Stegun, {\em Handbook of Mathematical Functions with
  Formulas, Graphs, and Mathematical Tables}.
\newblock Dover, New York, ninth dover printing, tenth gpo printing~ed., 1965.

\bibitem{Gibbons:2000hf}
G.~W. Gibbons, K.~Hori, and P.~Yi, {\it {String fluid from unstable D-branes}},
   {\em Nucl. Phys.} {\bf B596} (2001) 136--150,
  [\href{http://xxx.lanl.gov/abs/hep-th/0009061}{{\tt hep-th/0009061}}].

\bibitem{Gibbons:2002tv}
G.~Gibbons, K.~Hashimoto, and P.~Yi, {\it {Tachyon condensates, Carrollian
  contraction of Lorentz group, and fundamental strings}},  {\em JHEP} {\bf 09}
  (2002) 061, [\href{http://xxx.lanl.gov/abs/hep-th/0209034}{{\tt
  hep-th/0209034}}].

\bibitem{Sugimoto:2002fp}
S.~Sugimoto and S.~Terashima, {\it {Tachyon matter in boundary string field
  theory}},  {\em JHEP} {\bf 07} (2002) 025,
  [\href{http://xxx.lanl.gov/abs/hep-th/0205085}{{\tt hep-th/0205085}}].

\bibitem{Barnaby:2004nk}
N.~Barnaby, {\it {Caustic formation in tachyon effective field theories}},
  {\em JHEP} {\bf 07} (2004) 025,
  [\href{http://xxx.lanl.gov/abs/hep-th/0406120}{{\tt hep-th/0406120}}].

\bibitem{Sen:2002nu}
A.~Sen, {\it Rolling tachyon},  {\em JHEP} {\bf 04} (2002) 048,
  [\href{http://xxx.lanl.gov/abs/hep-th/0203211}{{\tt hep-th/0203211}}].

\bibitem{Larsen:2002wc}
F.~Larsen, A.~Naqvi, and S.~Terashima, {\it Rolling tachyons and decaying
  branes},  {\em JHEP} {\bf 02} (2003) 039,
  [\href{http://xxx.lanl.gov/abs/hep-th/0212248}{{\tt hep-th/0212248}}].

\bibitem{Copeland:2003df}
E.~J. Copeland, P.~M. Saffin, and D.~A. Steer, {\it Singular tachyon kinks from
  regular profiles},  {\em Phys. Rev.} {\bf D68} (2003) 065013,
  [\href{http://xxx.lanl.gov/abs/hep-th/0306294}{{\tt hep-th/0306294}}].

\end{thebibliography}\endgroup

\end{document}